\documentclass[lettersize,journal,twoside]{IEEEtran}
\usepackage{amsmath,amsfonts}
\usepackage{array}
\usepackage{url}
\usepackage{graphicx}
\usepackage{cite}
\usepackage{booktabs}
\usepackage{xcolor}
\usepackage{tikz}
\usepackage{algorithm}
\usepackage{algpseudocode}
\usepackage{multirow}
\usepackage{tabularx}
\begin{document}

\title{HHK: A Hardware-Oriented Cross-Location PPG Key Generation Architecture for Body Area Networks}

\author{Murilo~Calil~Faleiros and Jose~Ilton~de~Oliveira~Filho,~\IEEEmembership{Member,~IEEE}%
\thanks{M. Calil Faleiros is with King Abdullah University of Science and Technology (KAUST), Thuwal 23955-6900, Saudi Arabia (e-mail: murilo.calilfaleiros@kaust.edu.sa).}%
\thanks{J. I. d. O. Filho is with the Department of Computer and Information Engineering, Khalifa University, Abu Dhabi 127788, United Arab Emirates, and with King Abdullah University of Science and Technology (KAUST), Thuwal 23955-6900, Saudi Arabia (e-mail: joseilton.deoliveirafilho@ku.ac.ae).}%
\thanks{This research was supported by the Khalifa University Center for Autonomous Robotic Systems (KUCARS), Khalifa University (KU-KUCARS).}%
\thanks{Manuscript received XXX; revised XXX.}}

\markboth{}{}

\maketitle

\begin{abstract}
Body area networks (BANs) require session keys that wearable nodes establish without user interaction and with proof that both nodes are on the same body. This paper presents a multi-site photoplethysmography (PPG) key-agreement processor for BANs, incorporating signal conditioning, band-pass filtering, pulse onset detection, beat alignment, Gray-coded quantization and a polar-code fuzzy commitment. The algorithm reaches 95.6\% key agreement on resting wrist and fingertip recordings, and on 16 ambulatory participants the processor reaches 69.2, 54.8 and 50.0\% between head and wrist, wrist and ankle, and head and ankle, respectively, against 2.3\% for the best of five prior methods. Implemented and tested on an AMD XC7Z020 FPGA, the processor consumes 261\,\textmu W. Synthesized with the NanGate 45\,nm library, it occupies 0.394\,mm\textsuperscript{2}, operates at up to 258\,MHz and requires 0.208\,\textmu J per decoding. We further attack the system and show that the public helper data reveals more bits than the inter-beat intervals carry, so an eavesdropper who knows only the mean heart rate recovers 39.4\% of the session keys. To our knowledge, the processor is the first synthesizable RTL architecture for inter-beat-interval-based PPG key agreement in BANs evaluated on free-living ambulatory data.
\end{abstract}


\section{Introduction}

\begin{figure}[t]
\centering
\includegraphics[width=\columnwidth]{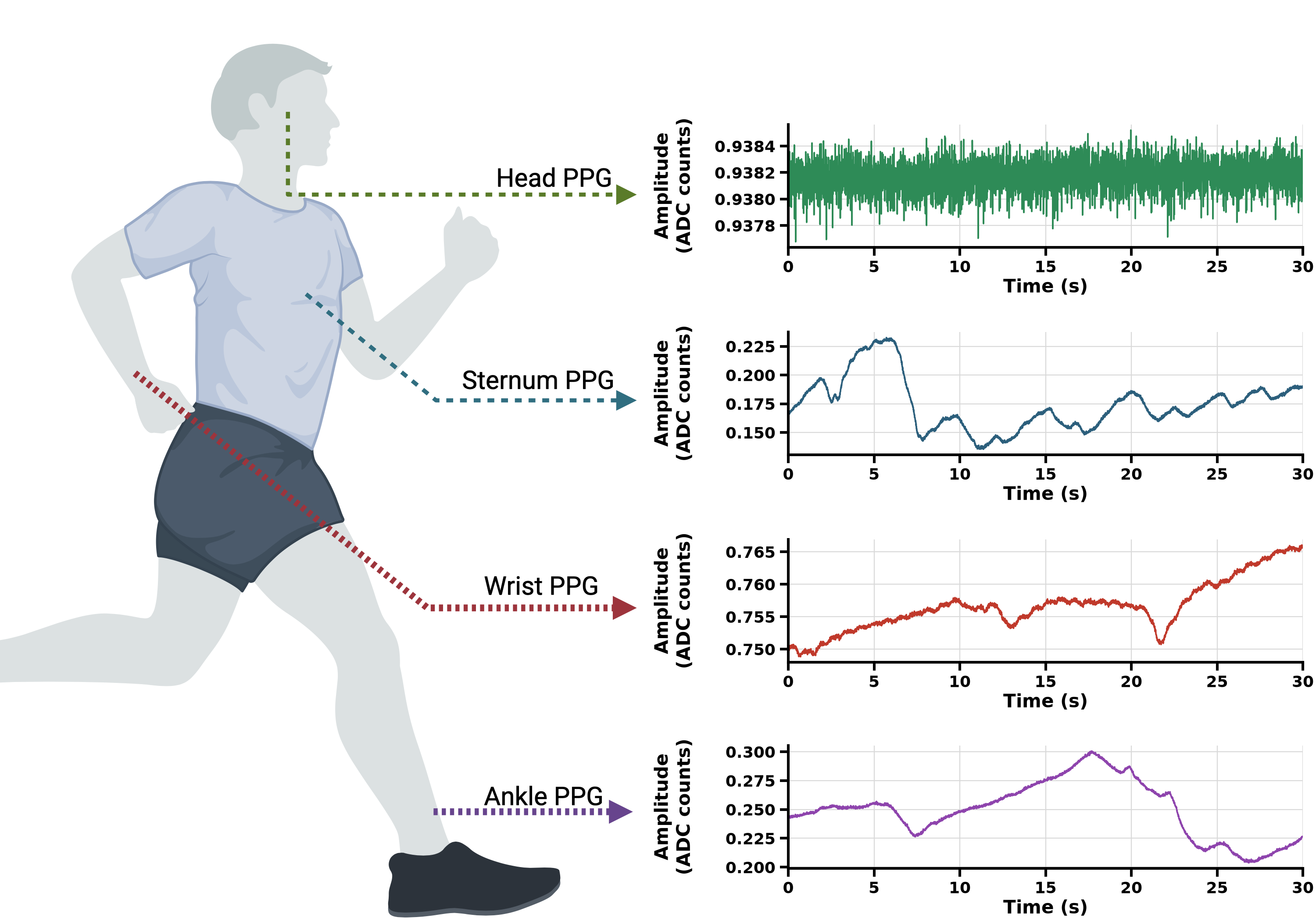}
\caption{Simultaneous PPG from an ambulatory participant (30\,s excerpt at 128\,Hz). Excerpt extracted from \cite{meier2024wildppg}.}
\label{fig:waveform}
\end{figure}

\IEEEPARstart{W}{ireless} body area networks (BANs) connect wearable and implantable sensing nodes for continuous health monitoring, sports tracking, and closed-loop therapy \cite{celik2022iobsurvey}. Unprotected BANs expose sensitive data to eavesdropping and command injection, and pairing every node by hand or with pre-distributed keys does not scale to networks of wearable devices. Physiological signal-based key generation pairs nodes with no user action and shows that both nodes are on the same body, because two nodes on the same body independently derive matching keys from shared physiology \cite{jiang2021fuzzy,guglielmi2022ecgkey}. Photoplethysmography (PPG) sensors measure the cardiac pulse at many body sites, and the inter-beat interval (IBI) sequence (called IPI in \cite{zhang2023h2k}) is common to all cardiac sensing modalities at rest \cite{benabdessalem2026crip}. Fig.~\ref{fig:waveform} illustrates the central cross-location challenge. Simultaneous PPG captures the same cardiac rhythm but differs in waveform shape and motion content.

The PPG key-generation studies closest to this work evaluate stationary participants, mostly at one location pair, so they leave open whether keys agree across sites during motion \cite{meier2024wildppg}. H2K \cite{zhang2023h2k} and the PPG token scheme \cite{mnijli2025ppgtoken} reach 94.1 and 96.6\% agreement at fingers and wrist, and FaceFinger \cite{wei2024facefinger} pairs fingertip and face video, all in laboratory recordings. E2P \cite{zhao2025e2p} and MagKey \cite{huang2025magkey} derive keys from ECG leads and wrist ballistocardiography in stationary recordings, so none of the five tests head, wrist and ankle together during free-living activity. These five systems, which this paper re-evaluates on ambulatory data, run entirely in software on a host processor, so their cost on a body node remains unmeasured. 

The proposed processor, HHK (Hardware Heart Key), derives the key from the IBI timing that all sites share, and this paper makes four contributions. (1)~We benchmark multi-location PPG key agreement on 216 hours of free-living data \cite{meier2024wildppg}, where the proposed algorithm reaches the highest key agreement of six methods in every location pair. (2)~We present a synthesizable RTL architecture of the complete key-generation pipeline, from the raw PPG samples to the reconciled message. To our knowledge, it is the first synthesizable RTL architecture for IBI-based PPG key agreement in BANs evaluated on free-living ambulatory data across three body locations. It is also the first PPG key-agreement hardware reported with post-implementation resource and power figures. (3)~We verify the RTL against a fixed-point model, on an FPGA board and in standard-cell netlists. (4)~We show with measured attacks that absolute-IBI quantization with a fuzzy commitment does not protect the key on ambulatory data, because the helper data exposes more bits than a block carries.

Section~\ref{sec:related} reviews the related work, Section~\ref{sec:arch} presents the key-agreement algorithm and Section~\ref{sec:exp} evaluates it on ambulatory and resting data. Section~\ref{sec:rtl} describes the RTL architecture, Section~\ref{sec:fpga} its implementation and verification, Section~\ref{sec:security} the security analysis and Section~\ref{sec:limits} the limitations, and Section~\ref{sec:concl} concludes the paper.

\section{Background and Related Work}
\label{sec:related}

Physiological key generation uses a signal that only sensors on the same body observe. Most schemes detect beats, quantize an IBI feature into bits at each node, and correct residual disagreements with public helper data \cite{jiang2021fuzzy,guglielmi2022ecgkey}. Cross-site heartbeat keys also have software precedents in F2H \cite{zheng2019f2h}, H2B \cite{lin2019h2b} and CBSP \cite{pourbemany2023cbsp}. PSKA \cite{venkatasubramanian2010pska} agrees symmetric keys between BAN sensors from physiological signals without predeployment, and H2H \cite{rostami2013h2h} authenticates a programmer to an implanted device from ECG inter-pulse intervals. Seepers~\emph{et al.} \cite{seepers2015vonneumann} measure 15.2\,bits of effective strength in a 60-bit inter-pulse-interval key. XKey \cite{zhao2026xkey} learns key features from ECG and gait with deep sequence models in software, and PPGTransID \cite{liu2026ppgbridge} authenticates wearables by correlating camera and contact PPG without generating a key.

 The two-node PPG key agreement of Banerjee~\emph{et al.} \cite{banerjee2009pka} targets a Spartan-II FPGA, and Karimian~\emph{et al.} \cite{karimian2019unlock} generate ECG keys on an AMD XC7Z020 without error correction. Yasin~\emph{et al.} \cite{yasin2017secureiot} extract a chip-specific ECG key in a 65\,nm ASIC flow, and Cherupally~\emph{et al.} \cite{cherupally2020securityengine} combine ECG and an SRAM physical unclonable function (PUF) in a 65\,nm chip.

Fuzzy commitment binds a random codeword of a code with block length $N$ and $K$ message bits to a noisy biometric string \cite{juels1999fuzzy}. Its helper data discloses at most $N-K$ bits about that string \cite{dodis2008fuzzy}. Polar codes achieve the capacity of binary-input symmetric memoryless channels \cite{arikan2009polarization}, and successive-cancellation (SC) decoding traverses the code tree once using log-likelihood ratios (LLRs). Semi-parallel decoders devote much of their area to LLR and partial-sum storage \cite{leroux2013semiparallel,fan2014partialsum}, and a combinational $N=128$ decoder uses 14\,517 lookup tables (LUTs) \cite{dizdar2016combinational}. The NIST SP 800-90B estimators bound the min-entropy of a quantized feature \cite{turan2018sp80090b}.

\section{Key-Agreement Algorithm}
\label{sec:arch}

\begin{figure}[t]
\centering
\includegraphics{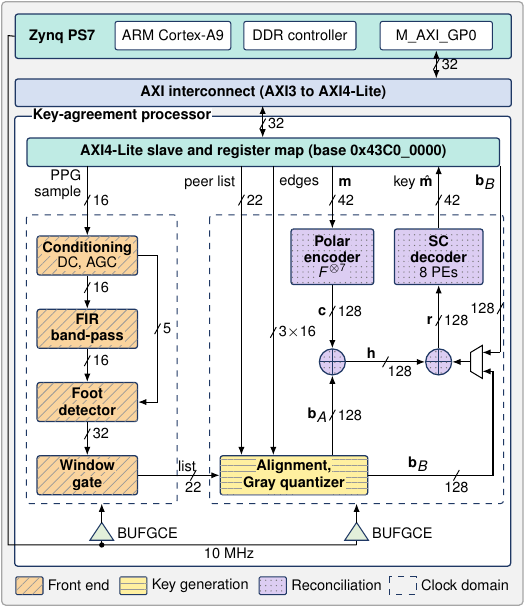}
\caption{Architecture of the processor. The $\oplus$ nodes form $\mathbf{h}=\mathbf{b}_A\oplus\mathbf{c}$ and $\mathbf{r}=\mathbf{b}_B\oplus\mathbf{h}$, and the dashed boxes mark the two gated clock domains.}
\label{fig:arch}
\end{figure}

Two BAN nodes at different body locations execute the same datapath (Fig.~\ref{fig:arch}) independently, and they exchange the beat timestamps that alignment needs and the helper data, while the message is never transmitted.

Raw 16-bit PPG samples are DC-removed, scaled by a power-of-two gain and band-pass filtered between 0.5 and 4\,Hz. A detector then marks the pulse foot of each cycle by tangent intersection. Let $t_u(n)$ denote the $n$th beat timestamp at body site~$u$. The inter-beat interval is
\begin{equation}
 \mathrm{IBI}_u(n) = t_u(n+1) - t_u(n).
 \label{eq:ibi}
\end{equation}

Algorithm~\ref{alg:ibi} pairs each site-$A$ IBI with the nearest site-$B$ beats within a 300\,ms tolerance, and only the kept pairs contribute to IBI sequences, which removes the errors that missed or extra beats cause when IBIs are compared in sequence. The IBI range is partitioned into four equal-frequency bins using dataset-wide percentiles, and each IBI is mapped to a 2-bit Gray code. Gray coding limits a boundary crossing to a 1-bit disagreement, which reduces the effective channel error rate.

\begin{algorithm}[t]
\caption{IBI Extraction and Cross-Location Alignment}
\label{alg:ibi}
\small
\begin{algorithmic}[1]
\Require PPG streams $x_A(t)$, $x_B(t)$ at $f_s=128$\,Hz, tolerance $\delta=300$\,ms
\Ensure Aligned IBI pairs and Gray-coded bit strings $\mathbf{b}_A,\mathbf{b}_B\in\{0,1\}^{128}$
\For{each site $u \in \{A,B\}$}
 \State Band-pass filter $x_u(t)$
 \State Detect one pulse foot $t_u(n)$ per cycle
 \State Compute $\mathrm{IBI}_u(n)=t_u(n+1)-t_u(n)$
\EndFor
\State $\mathcal{M}\leftarrow\emptyset$
\For{each $i\geq1$}
 \State $j_0=\arg\min_j|t_A(i-1)-t_B(j)|$, $j_1=\arg\min_j|t_A(i)-t_B(j)|$
 \If{$|t_A(i-1)-t_B(j_0)|\leq\delta$, $|t_A(i)-t_B(j_1)|\leq\delta$, $j_0\neq j_1$}
 \State $\mathcal{M}\leftarrow\mathcal{M}\cup\{(t_A(i)-t_A(i-1),\,t_B(j_1)-t_B(j_0))\}$
 \EndIf
\EndFor
\State Discard pairs outside 400 to 1\,333\,ms or beyond $2\sigma$ of the window mean
\State Map each IBI in $\mathcal{M}$ to a 2-bit Gray code with the equal-frequency bin edges and concatenate the codes into $\mathbf{b}_A,\mathbf{b}_B$
\end{algorithmic}
\end{algorithm}

Node $A$ draws a random message $\mathbf{m} \in \{0,1\}^{42}$, encodes it to codeword $\mathbf{c}$ under a rate-1/3 polar code ($N=128$, $K=42$), and broadcasts helper data $\mathbf{h} = \mathbf{b}_A \oplus \mathbf{c}$. Node $B$ computes $\mathbf{r} = \mathbf{b}_B \oplus \mathbf{h}$ and decodes it with the SC decoder to recover $\hat{\mathbf{m}}$. Reconciliation succeeds when $\hat{\mathbf{m}} = \mathbf{m}$, and a failed window triggers a retry with the next window. Three successful blocks yield 126 bits of shared key material, which SHA-256 hashes into a 128-bit session key. We chose $N=128$ because a 120\,s window yields 105 to 110 aligned IBIs, that is 210 to 220 bits, and 128 is the largest power of two below this count. We chose $K=42$ within the capacity bound $128[1-H_2(\bar p)]$, which admits any $K\leq54$ at the pooled head-ankle mismatch of 13.7\%.

\section{Experimental Validation}
\label{sec:exp}

\subsection{Dataset and Protocol}
\label{sec:protocol}

We use WildPPG \cite{meier2024wildppg}, a 2024 dataset recorded during a day trip to the Jungfraujoch, Switzerland, at 3\,460\,m. Sixteen participants wore synchronized PPG and inertial measurement unit (IMU) sensors at four body locations for an average of 13.5 hours each, totaling 216 recording hours. We use the head, wrist and ankle PPG channels.

We evaluate with 120\,s windows, and each location is gated for motion independently by its IMU standard deviation before cross-location matching. Consecutive windows start 60\,s apart, and a location passes when the standard deviation of its accelerometer magnitude is at most 0.035\,g. Windows with a heart rate outside 45 to 120\,BPM or an IBI coefficient of variation above 20\% are discarded, and equal-frequency bin edges are computed from all retained IBIs across the 16 participants. A window counts only when the location pair yields at least 64 aligned IBIs (128 bits). Each window draws a fresh random message $\mathbf{m}$, and key agreement uses the fixed frozen set of the RTL (Section~\ref{sec:rtl_enc}).

\subsection{Results}
\label{sec:results}

\begin{table*}[t!]
\centering
\caption{Comparison With Prior Physiological Key Generation Work}
\label{tab:comparison}
\renewcommand{\arraystretch}{0.98}
\setlength{\tabcolsep}{2.5pt}
\footnotesize

\begin{tabularx}{\textwidth}{@{}l l c X X c c c@{}}
\toprule
Paper & Method & Cond. & Dataset & Locations & Orig. Agr. (\%) & Raw Bit (\%) & WildPPG Agr. (\%) \\
\midrule

H2K \cite{zhang2023h2k}
& IBI+BCH (ECG/PPG)
& Lab
& MIT-BIH/Testbed
& Finger-Finger
& 94.1
& 51.9
& 0.2 \\

FaceFinger \cite{wei2024facefinger}
& IBI+CS (PPG)
& Lab
& Testbed
& Finger-Face
& $\leq$100
& 58.9
& 0.5 \\

E2P \cite{zhao2025e2p}
& Ampl.+Cascade (ECG)
& Lab
& LUDB/LSDB/SPM
& ECG leads I, V4 to V6
& $\approx$100
& 52.0
& 2.2 \\

MagKey \cite{huang2025magkey}
& PPT+FC (BCG)
& Lab
& Custom (30 part.)
& Wrist-Wrist
& 94.2
& 53.2
& 0.0 \\

PPG Token \cite{mnijli2025ppgtoken}
& IBI (PPG)
& Lab
& Custom (7 part.)
& Finger, Wrist
& 96.6
& 61.6
& 0.0 \\

\multirow[t]{3}{*}{\textbf{This work}}
& \multirow[t]{3}{*}{\textbf{IBI+polar (PPG)}}
& \multirow[t]{3}{*}{\textbf{Amb.}}
& \multirow[t]{3}{*}{\textbf{WildPPG \cite{meier2024wildppg}}}
& \textbf{Head-Wrist}
& \textbf{n/a}
& \textbf{90.3}
& \textbf{69.3} \\

& & & & \textbf{Head-Ankle}
& \textbf{n/a}
& \textbf{86.3}
& \textbf{50.4} \\

& & & & \textbf{Wrist-Ankle}
& \textbf{n/a}
& \textbf{87.8}
& \textbf{55.2} \\

\bottomrule
\end{tabularx}
\end{table*}

The aligned IBIs of the two sites differ by 25.5 to 31.1\,ms on average (mean absolute error, MAE). Per reconciliation window, polar decoding corrects the residual disagreements and yields 69.3\% key agreement between head and wrist at 90.3\% raw agreement, 55.2\% between wrist and ankle at 87.8\% and 50.4\% between head and ankle at 86.3\%. With non-overlapping retries, head-wrist needs 1.44 attempts on average and head-ankle 1.98, that is 173 to 238\,s of usable signal per reconciled block. The motion gate passes 10.9 to 21.8\% of the windows, so a node completes 0.20 to 0.38 session keys per hour of wear. In Table~\ref{tab:comparison}, Cond.\ gives the recording condition (Lab for controlled and stationary, Amb.\ for free-living ambulatory), Orig.\ Agr.\ gives the agreement each paper reports on its own data, and n/a marks a value that does not apply. For H2K and PPG Token, Orig.\ Agr.\ is 100\% minus the reported key disagreement rate.

Re-evaluated on WildPPG with their own gating (0.08\,g, IBI CV at most 30\%), the competing methods reach 52 to 62\% raw bit agreement and 0.0 to 2.2\% key agreement on all three pairs (head-wrist values in the Raw Bit and WildPPG Agr.\ columns of Table~\ref{tab:comparison}). Under the same gating, the algorithm reaches 78.2 to 80.8\% raw and 40.7 to 55.7\% key agreement. Under the stricter gate of this work, the competing methods reach 0.0 to 2.3\%, so the comparison does not depend on the gate. With the aligned IBI pairs of HHK as input and under its gate, H2K and FaceFinger reach 20.2 and 26.3\% head-wrist key agreement. We re-implemented each method from its paper, with one permutation shared by both E2P nodes. H2K and MagKey count a block as reconciled when its mismatches stay within the capability of their codes, and FaceFinger corrects with both bit strings known, which favors it.

The design choices of Section~\ref{sec:arch} hold between 200 and 400\,ms of alignment tolerance, where the raw agreement changes by 1.6, 1.2 and 1.0\,pp and the key agreement by 3.3, 2.5 and 3.6\,pp for head-wrist, head-ankle and wrist-ankle. Below 200\,ms the ankle pairs lose paired beats. Equal-frequency bins carry 1.95\,bits per symbol and uniform-width bins 1.29\,bits. Uniform-width bins raise the key agreement to 85.6, 77.2 and 77.0\% for head-wrist, head-ankle and wrist-ankle, so equal-frequency binning gives up agreement for higher entropy.

\subsection{Evaluation on Resting Datasets}
\label{sec:rest}

The stationary studies of Table~\ref{tab:comparison} report resting agreement, so we ran the algorithm on two resting datasets. WESAD \cite{schmidt2018wesad} records 15 participants with chest ECG and wrist PPG through baseline, amusement, stress and meditation sessions. Node A takes its beats from the ECG (Pan and Tompkins \cite{pan1985qrs}) and node B from the PPG resampled to 128\,Hz, so the pair tests IBI agreement across modalities. The ECG and PPG beat sequences of the released files drift by $-0.39$ to $+3.53$\,s, so we remove a per-window offset, and without it the MAE grows to 59.1\,ms and the per-window key agreement falls to 22.3\%. 

With the fixed frozen set and bin edges derived on each dataset, the algorithm reaches 85.5\% key agreement on WESAD and 95.6\% on WF-PPG \cite{ho2025wfppg}, which has 27 seated participants with PPG on the wrist and the fingertip of the same hand. The stationary studies report 94.1 to 100\% on their own data, as bit agreement or key success. WF-PPG measures two sites of one hand at rest, while WildPPG adds motion and larger site distances, so this comparison does not separate the two effects.

\section{RTL Architecture}
\label{sec:rtl}

\subsection{System Integration and Data Flow}
\label{sec:rtl_sys}

Behind an AXI4-Lite slave, the processor (Fig.~\ref{fig:arch}) holds a front end (conditioning stage, filter, foot detector and window gate) and a key-generation unit (alignment unit with the Gray quantizer), and a reconciliation back end (polar encoder and SC decoder) completes it. The design targets ASIC integration in a BAN node, and the FPGA prototype verifies the RTL stage by stage against the software. No module uses a DSP slice or block RAM, and every variable multiplication uses a sequential shift-and-add unit and every division a restoring divider.

A host write to the sample register produces a one-cycle strobe at 128\,Hz. The sample-rate modules then need 4\,379 cycles per sample on average, 5.6\% of the 78\,125 cycles of a 10\,MHz clock. Alignment, quantization, encoding and decoding then take 62\,223 to 175\,054 cycles (6.2 to 17.5\,ms).

Two BUFGCE clock gates on the 10\,MHz clock of the prototype give a front-end domain and a domain for the key-generation unit and the reconciliation, which receive 560\,560 and 967 edges per second over a window. The prototype implements one node and both roles of the protocol, and the host writes the beat list of the other site as the peer list. The host also supplies the random message and the motion gate, and hashes the session key. The alignment unit outputs both bit strings, and a host register can replace $\mathbf{b}_B$ for tests. 

\subsection{Architecture of Each Module}
\label{sec:modules}

\subsubsection{Conditioning}
\label{sec:rtl_cond}
The stage removes the DC level of the raw code $x[n]$ with the tracker $d[n]=d[n-1]+2^{-10}(x[n]-d[n-1])$ (8\,s time constant), whose 28-bit accumulator starts from the first sample. It shifts the high-pass value $x[n]-d[n-1]$ by a gain $2^{g}$, $g\in\{-3,\ldots,12\}$, and saturates the result to the 16-bit filter input $c[n]$. Every 32\,s ($2^{12}$ samples) it compares the block peak with fixed thresholds, and a new gain takes effect at the next zero crossing, where the step into the filter is smallest. The stage uses 280 LUTs and 87 flip-flops (FFs).

\subsubsection{Band-Pass FIR Filter}
\label{sec:rtl_bpf}
\begin{figure}[t]
\centering
\includegraphics{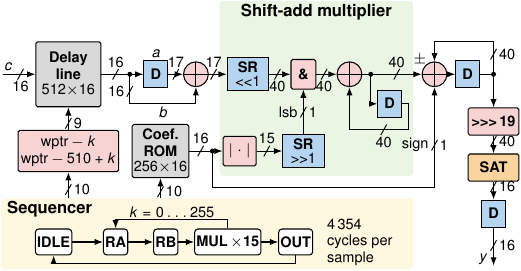}
\caption{Architecture of the band-pass FIR filter module. The pointer wptr is the write address of the delay line.}
\label{fig:bpf}
\end{figure}

The filter (Fig.~\ref{fig:bpf}) is a 511-tap linear-phase FIR whose integer taps $h_k$ are the impulse response of the zero-phase software filter, truncated to $\pm2$\,s with a Hann taper and adjusted to zero DC gain,
\begin{equation}
\begin{aligned}
y[n]=\mathrm{sat}_{16}\Big(2^{-19}\Big[&\sum_{k=0}^{254}h_k\big(c[n-k]+c[n-510+k]\big)\\
&+h_{255}\,c[n-255]\Big]\Big).
\end{aligned}
\label{eq:fir}
\end{equation}
A $512\times16$ LUT-RAM holds the delay line, which is read at $\mathrm{wptr}-k$ and $\mathrm{wptr}-510+k$, so the symmetric taps share one pre-addition. One shift-and-add multiplier with a $256\times16$ coefficient ROM feeds a 40-bit accumulator in $(L+1)\cdot17+2=4\,354$ cycles per output, with $L=255$. On 288 training windows, causal Butterworth filters shifted the detected feet by 17 to 29\,ms on average, and linear-phase versions by 0.9 to 4.9\,ms. The filter uses 576 LUTs (176 as LUT-RAM) and 213 FFs.

\subsubsection{Foot Detector}
\label{sec:rtl_foot}
\begin{figure}[t]
\centering
\includegraphics{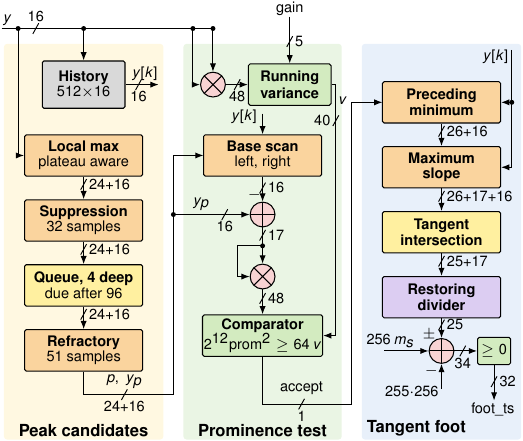}
\caption{Architecture of the foot detector module. The $y[k]$ inputs are read from the $512\times16$ history RAM.}
\label{fig:foot}
\end{figure}

The detector (Fig.~\ref{fig:foot}) keeps the last 512 filtered samples in LUT-RAM and applies the software peak rules to them. A plateau-aware local maximum is suppressed when a higher one lies within 32 samples (250\,ms). A remaining maximum is accepted when its prominence prom satisfies $2^{12}\,\mathrm{prom}^2\geq64\,v$, with $v\leftarrow v+2^{-10}(2^4y^2-v)$ a running variance that follows the gain steps. This test equals $\mathrm{prom}\geq\sigma/2$ for the standard deviation $\sigma$ of $y$. The foot is the intersection of the tangent at the steepest preceding point $m_s$ with the level of the preceding minimum. A 25-iteration restoring divider gives its sub-sample position, and timestamps are output in q8 format (1/256 sample, 30.5\,\textmu s) with the 255-sample filter delay ($255\cdot256$ q8 units) removed. The detector uses 2\,196 LUTs (176 as LUT-RAM) and 940 FFs.

\subsubsection{Window Gate}
\label{sec:rtl_gate}
\begin{figure}[t]
\centering
\includegraphics{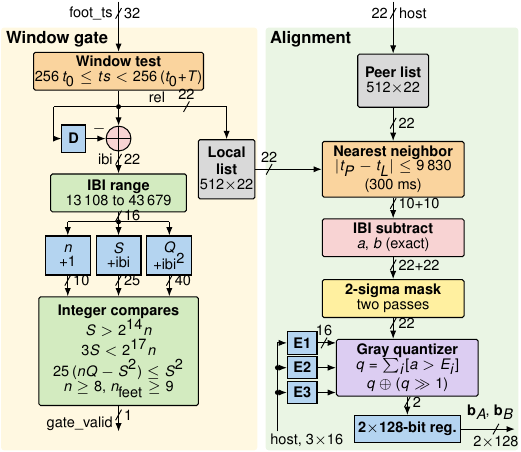}
\caption{Architecture of the window gate and alignment unit modules.}
\label{fig:gatealign}
\end{figure}

The gate (Fig.~\ref{fig:gatealign}) keeps the feet with $256\,t_0\leq ts<256\,(t_0+T)$, with $t_0$ the first sample and $T$ the length of the window, and stores them relative to $t_0$ in a $512\times22$ LUT-RAM that also serves as the local list of the alignment unit. It accumulates the number $n$, the sum $S$ and the sum of squares $Q$ (10, 25 and 40 bits) of the IBIs between 13\,108 and 43\,679 q8 units (400 and 1\,333\,ms). It accepts the window when $n\geq8$, at least 9 feet are stored and
\begin{equation}
S>2^{14}n,\qquad 3S<2^{17}n,\qquad 25\,(nQ-S^2)\leq S^2,
\label{eq:gate}
\end{equation}
which implement $45<\mathrm{HR}<120$\,BPM and $\mathrm{CV}\leq20\%$ without a division. The motion gate needs the accelerometer and stays on the host. The gate uses 1\,035 LUTs (240 as LUT-RAM) and 319 FFs.

\subsubsection{Alignment Unit and Gray Quantizer}
\label{sec:rtl_align}
The unit runs Algorithm~\ref{alg:ibi} on the local foot list and on a peer list that the host writes into a second $512\times22$ LUT-RAM. For every local beat it scans the peer list for the nearest beat and keeps the pair when both ends lie within 9\,830 q8 units (300\,ms) and map to distinct peer beats. It then forms the IBIs $a$ and $b$ of both sites from the unrounded timestamps and drops pairs outside the IBI range. A second pass keeps a pair when $(na-S)^2<4(nQ-S^2)$ at both sites, the integer $2\sigma$ test. Three edge registers then give the bin $q=[a>E_1]+[a>E_2]+[a>E_3]$ and the Gray symbol $q\oplus(q\gg1)$. The unit uses 2\,254 LUTs (240 as LUT-RAM) and 1\,176 FFs.

\subsubsection{Polar Encoder}
\label{sec:rtl_enc}
The encoder writes the 42 message bits into the information set
\begin{equation}
\begin{aligned}
\mathcal{A}=\{&55,\,59{:}63,\,79,\,86,\,87,\,89{:}95,\,99,\\
&101{:}103,\,105{:}111,\,113{:}127\}
\end{aligned}
\label{eq:infoset}
\end{equation}
and zeros the 86 frozen positions, with $a{:}b$ the integers from $a$ to $b$. The set is the Bhattacharyya construction for a crossover probability of 0.10, fixed in the RTL. Seven combinational stages of 64 XOR gates compute $\mathbf{u}F^{\otimes7}$. Encoder and control use 214 LUTs and 375 FFs.

\subsubsection{Semi-Parallel SC Decoder}
\label{sec:rtl_dec}
\begin{figure}[t]
\centering
\includegraphics{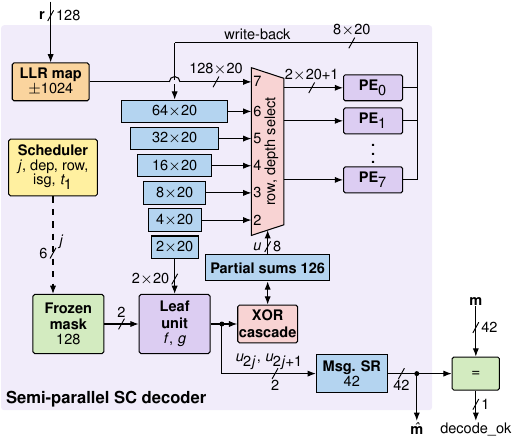}
\caption{Architecture of the SC decoder module. The comparison with $\mathbf{m}$ that gives \texttt{decode\_ok} belongs to the core and is drawn outside the decoder.}
\label{fig:decoder}
\end{figure}

The decoder (Fig.~\ref{fig:decoder}) maps each bit $r_i$ to an LLR of $\pm1\,024$ and traverses the depth-7 code tree. A node computes the left-child LLRs with $f(a,b)=\mathrm{sgn}(a)\,\mathrm{sgn}(b)\min(|a|,|b|)$ and the right-child LLRs with $g(a,b,\hat{c})=b+(1-2\hat{c})a$, with $\hat{c}$ the partial sum of the left child. One LLR vector per depth holds the state, with 64, 32, 16, 8, 4 and 2 words of 20 bits for depths 6 to 1, and no channel LLR is stored. Eight processing elements (PEs) give $\max(1,2^{d-4})$ cycles per step at depth $d$, and a leaf unit resolves one depth-1 node per cycle. A 126-bit register set and an XOR cascade keep the partial sums \cite{leroux2013semiparallel}, and a 42-bit shift register collects $\hat{\mathbf{m}}$. With $P$ PEs and the start cycle, one decoding takes
\begin{equation}
C(P)=65+\sum_{d=2}^{7}2^{8-d}\max\!\left(1,\frac{2^{d-1}}{P}\right)
\label{eq:cycles}
\end{equation}
cycles, $C(8)=225$ or 22.5\,\textmu s at 10\,MHz. Every LLR is a multiple of 1\,024 bounded by $2^{17}$, so the words cannot overflow. The decoder uses 1\,802 LUTs and 2\,759 FFs, and the PE count trades area against cycles.

\section{FPGA Implementation and Verification}
\label{sec:fpga}

\subsection{Verification Against the Fixed-Point Model}
\label{sec:modelval}

Self-checking testbenches compare every module output with a bit-exact fixed-point model of the RTL (hereafter the fixed-point model) on WildPPG segments and boundary cases, and they detected all 29 faults inserted by hand. Vivado xsim of two complete nodes reproduced the fixed-point model in all 25 windows of 8 participants, in every foot, gate decision and aligned pair and in $\mathbf{b}_A$, $\mathbf{b}_B$, $\mathbf{h}$, $\mathbf{r}$ and $\hat{\mathbf{m}}$. The decoder reproduced the decisions of a reference SC decoder in RTL on all 11\,208 received words, and the PE with the kernels $f$ and $g$ on all 66\,049 pairs of reachable LLRs.

On all 16 participants, the mean absolute difference between the feet of the model and of the software is 2.0, 2.3 and 2.6\,ms at the head, wrist and ankle, and the model finds 96.3, 95.7 and 94.0\% of the software beats. For head-wrist, head-ankle and wrist-ankle it reaches 90.6, 86.5 and 88.2\% raw and 69.2, 50.0 and 54.8\% key agreement, against 90.3, 86.3 and 87.8\% and 70.0, 50.7 and 55.8\% for the software with the same messages. Over five message seeds, the key agreement of the software varies with a standard deviation of 0.31, 0.51 and 0.69\,pp. We tuned the filter length, peak distance, decision delay, refractory period, prominence factor, variance time constant and gain control on 8 participants, while the bin edges come from all 16 (Section~\ref{sec:protocol}). On the other 8 participants the model is 2.9, 1.4 and 0.7\,pp below the software. Bin edges from the 8 training participants alone give 61.1, 56.3 and 60.4\% head-wrist, head-ankle and wrist-ankle key agreement on the held-out participants, against 58.6, 55.3 and 56.5\% with the population edges. The population edges therefore do not inflate the held-out result.

\subsection{Implementation Results}
\label{sec:synth}
\label{sec:asic}

Post-implementation results on an AMD XC7Z020 (Pynq-Z2 board, Vivado~2025.2, 10\,MHz), give 8\,417 LUTs and 6\,259 FFs for the processor (Table~\ref{tab:modules}). The system with the AXI interconnect uses 8\,799 LUTs and 6\,722 FFs, 16.5\% of the LUTs and 6.3\% of the FFs of the device. The only memory primitives are the 832 LUT-RAM cells of the four 512-word arrays. The foot detector and the alignment unit are the largest modules. Timing closes at 10\,MHz with 72.3\,ns of setup slack, and in an out-of-context run the processor meets a 16\,ns period and fails at 14\,ns (62.5 to 71.4\,MHz).

\begin{table}[t]
\centering
\caption{Post-Implementation Resource Utilization and Power (AMD XC7Z020, Vivado~2025.2, 10\,MHz)}
\label{tab:modules}

\footnotesize
\setlength{\tabcolsep}{4pt}
\begin{tabular}{@{}lrrr@{}}
\toprule
Module & LUTs & LUT-RAM & FFs \\
\midrule
Conditioning & 280 & 0 & 87 \\
Band-pass FIR, 511 taps & 576 & 176 & 213 \\
Foot detector & 2\,196 & 176 & 940 \\
Window gate & 1\,035 & 240 & 319 \\
Alignment unit, Gray quantizer & 2\,254 & 240 & 1\,176 \\
\quad Front end and key generation & 6\,341 & 832 & 2\,735 \\
Polar encoder and control & 214 & 0 & 375 \\
SC decoder, 8 PEs & 1\,802 & 0 & 2\,759 \\
AXI4-Lite slave & 87 & 0 & 390 \\
\midrule
Processor total & 8\,417 & 832 & 6\,259 \\
System total (with AXI interconnect) & 8\,799 & 832 & 6\,722 \\
\midrule
Dynamic power (\textmu W) & \multicolumn{3}{r}{261} \\
Energy per 120\,s window (mJ) & \multicolumn{3}{r}{31.3} \\
Cycles per decoding & \multicolumn{3}{r}{225} \\
\bottomrule
\multicolumn{4}{@{}p{\columnwidth}@{}}{\scriptsize The module rows sum to 8\,444 LUTs, 27 more than the total, because Vivado packs logic of adjacent modules into shared LUTs.} \\
\end{tabular}
\end{table}

We compute the dynamic power of the routed design in Vivado from simulated switching activity. Vivado models a gated clock at its full frequency, so we constrain each gated clock at its simulated average edge rate. Every other net takes its simulated toggle rate, except the AXI nets at 256 transitions per second. The processor then draws 261\,\textmu W, or 31.3\,mJ per 120\,s window, of which the free-running AXI4-Lite slave accounts for 196\,\textmu W and the front end for 62\,\textmu W. At the key rate of Section~\ref{sec:results}, the processor consumes 2.5 to 4.7\,J per session key. Without the free-running AXI4-Lite slave of the prototype, which a body node does not need, it draws 65\,\textmu W, 4.0 times less.

Because the design uses no hard IP, it ports to standard cells. We synthesized the RTL with Yosys~0.67 and ABC to the NanGate 45\,nm library (typical, 25\,$^\circ$C, 1.10\,V). A library latch and an AND gate model each BUFGCE, and OpenSTA~3.1 analyzed the netlists with wire-load models. The core occupies 0.394\,mm$^2$ and operates at up to 258\,MHz. Of the 0.365\,mm$^2$ of the front end with the key-generation unit, 0.176\,mm$^2$ are the 38\,912 FFs of the four 512-word arrays, because the NanGate library has no SRAM macros. 

In zero-delay gate-level simulation of a full window, the core netlist reproduced the RTL in every output and in a 64-bit signature updated at every cycle. One decoding dissipates 9.27\,mW for 22.5\,\textmu s (0.208\,\textmu J), of which 7.52\,mW is leakage. 

Table~\ref{tab:hwcompare} compares this work with biosignal-security, biosignal-processing and key-reconciliation hardware, where ECC marks error correction or reconciliation in hardware and Silicon a measured chip. In the table, n/r marks a value that the work does not report, n/a one that does not apply, and all competitor power values are for continuous operation. The table leaves out the five software systems of Table~\ref{tab:comparison}, which report no hardware. It also leaves out the 22\,nm PUF decoders \cite{kestel2022polarpuf,wang2026polarpuf}, because the $N=128$ decoder of Dizdar and Ar{\i}kan already represents decoder-only hardware. Power is the metric where this work is worse, at 261\,\textmu W against 4.04 to 62.37\,\textmu W for four of the seven biosignal circuits, because of the free-running AXI4-Lite slave.

\begin{table*}[t]
\centering
\caption{Comparison With Biosignal-Security, Biosignal-Processing, and Key-Reconciliation Hardware}
\label{tab:hwcompare}
\scriptsize
\renewcommand{\arraystretch}{1.05}
\setlength{\tabcolsep}{2.6pt}
\begin{tabularx}{\textwidth}{@{}>{\raggedright\arraybackslash}p{1.85cm} >{\raggedright\arraybackslash}X >{\raggedright\arraybackslash}p{1.5cm} >{\raggedright\arraybackslash}p{1.42cm} >{\raggedright\arraybackslash}p{1.8cm} >{\raggedright\arraybackslash}p{1.35cm} >{\raggedright\arraybackslash}p{1.45cm} >{\raggedright\arraybackslash}p{1.75cm} >{\raggedright\arraybackslash}p{0.7cm} >{\raggedright\arraybackslash}p{1.05cm}@{}}
\toprule
Work & Function (signal) & Technology & Status & Area & Supply, clock & Power & Energy, latency & Key agr. & On-chip ECC \\
\midrule
Banerjee \textit{et al.}~\cite{banerjee2009pka}
 & Two-node PPG key agreement, fuzzy vault (PPG, two fingers, seated)
 & Spartan-II FPGA & VHDL sim.$^{(d)}$
 & n/r (47.35/45.3\,KB memory)
 & n/r, 20\,MHz & n/r & 32.2/59\,ms latency
 & Yes & Yes (vault) \\
Cherupally \textit{et al.}~\cite{cherupally2020ecgauth}
 & ECG authentication, sparse NN (ECG)
 & 65\,nm CMOS & Post-layout$^{(b)}$
 & 3.06\,mm$^2$ (I/O-limited)
 & 1.2\,V, 10\,kHz & 62.37\,\textmu W & 1.25\,mJ per auth.$^{(a)}$
 & No & No \\
Cherupally \textit{et al.}~\cite{cherupally2020securityengine}
 & ECG authentication and single-device key, fuzzy extractor off-chip (ECG)
 & 65\,nm CMOS & Silicon
 & 7.54\,mm$^2$ chip$^{(c)}$
 & 0.6\,V, 2.2\,kHz & 4.04\,\textmu W & 121\,\textmu J per auth.$^{(a)}$
 & No & No \\
Karimian \textit{et al.}~\cite{karimian2019unlock}
 & Single-device ECG key generation, no ECC implemented (ECG)
 & AMD XC7Z020 & FPGA
 & 2\,232 LUT, 1\,064 FF, 21 DSP
 & n/r & 39\,mW & n/r
 & No & No \\
Chang \textit{et al.}~\cite{chang2023ppgprocessor}
 & PPG identification and HR estimation (PPG, ACC)
 & 180\,nm CMOS & Silicon
 & 5.80\,mm$^2$, 190\,kGE
 & 1\,V, 320\,kHz & 427\,\textmu W & 17.1\,\textmu J/ID$^{(a)}$
 & No & No \\
Lee \textit{et al.}~\cite{lee2018hbll}
 & PPG HR sensor, heartbeat-locked loop (PPG)
 & 0.18\,\textmu m CMOS & Silicon
 & 1.51\,mm$^2$ die$^{(a)}$
 & 3.3\,V, 40 to 100\,Hz & 43.4\,\textmu W incl.\ LED & 43.4\,\textmu J/beat$^{(a)}$
 & No & No \\
Wang \textit{et al.}~\cite{wang2024trimodal}
 & HR, SpO$_2$ and pulse transit time interface (PPG)
 & 65\,nm CMOS & Silicon
 & 1.52\,mm$^2$
 & 0.6/1.2\,V, 100\,kHz & 18.9 to 43.7\,\textmu W & n/r
 & No & No \\
Dizdar and Ar{\i}kan~\cite{dizdar2016combinational}
 & Combinational SC polar decoder, $N{=}128$ (none)
 & Virtex-6, 90\,nm & P\&R, synth.
 & 14\,517 LUT, 783 FF, 0.338\,mm$^2$
 & 1.3\,V, 22.2\,MHz & 138.8\,mW (ASIC) & 49.0\,pJ/bit (ASIC)
 & n/a & Decoder only \\
\midrule
\textbf{This work, FPGA}
 & \textbf{\boldmath Cross-location IBI key agreement from raw samples to a polar $(128,42)$ fuzzy commitment with SC decoder (PPG)}
 & \textbf{AMD XC7Z020} & \textbf{FPGA board run}
 & \textbf{8\,417 LUT (832 LUT-RAM), 6\,259 FF, 0 DSP, 0 BRAM}
 & \textbf{1.0\,V core, 10\,MHz} & \textbf{261\,\textmu W} & \textbf{31.3\,mJ per 120\,s window, 22.5\,\textmu s per decode}
 & \textbf{Yes} & \textbf{Yes (polar SC)} \\
\textbf{This work, standard cells}
 & \textbf{Same RTL (Section~\ref{sec:asic})}
 & \textbf{NanGate 45\,nm} & \textbf{Post-synth.}
 & \textbf{\boldmath 0.394\,mm$^2$, 494\,kGE}
 & \textbf{1.10\,V, 10\,MHz} & \textbf{9.27\,mW decoding (7.52\,mW leakage)} & \textbf{0.208\,\textmu J per decode}
 & \textbf{Yes} & \textbf{Yes (polar SC)} \\
\bottomrule
\multicolumn{10}{@{}p{\textwidth}@{}}{(a)~Derived by us. (b)~Not fabricated. (c)~Die photograph, 5.54\,mm$^2$. (d)~From cycle counts, without synthesis.} \\
\end{tabularx}
\end{table*}

\subsection{Board Verification}
\label{sec:hwval}

On the board, a Python script on the processing system drives the processor through its registers only. For each window it resets the pipeline, loads the edges, streams the raw samples of both sites, writes the foot list of the second site as the peer list and logs every checkpoint with a timestamp. We ran 14 windows from 4 participants covering the three pairs, 13 with the recorded second site and one with a constant host-written $\mathbf{b}_B$. The board reproduced xsim of the RTL in all 7\,121 checkpoint reads.

The processor found 3\,225 beats against 3\,236 software feet, and 96.9\% of the software feet had a processor foot within 150\,ms, at a median offset of $-1$ to $+1$\,ms. With the recorded peer, 9 of 13 windows reconciled, and the mean raw agreement of the 13 windows was 91.5\%.

\section{Security Analysis}
\label{sec:security}

We evaluate the algorithm on the same 16 participants, windows and gating as Table~\ref{tab:comparison}, with the fixed frozen set of the RTL decoder. The experiments draw their own messages, which give the genuine key agreement of Table~\ref{tab:sec_leakage}. The RTL quantizes with the same edges and alignment rule, so the analysis applies to it.

\subsection{Adversarial Model}
\label{sec:sec_model}

Two nodes on one body run the protocol of Section~\ref{sec:arch} over a public channel. The adversary knows the protocol, the code with its frozen set, the quantizer edges and the alignment rule. The secrets are $\mathbf{b}_A$ and the message $\mathbf{m}$ of each block, and the session key is the SHA-256 digest of three messages. The adversary wins when it learns a session key, or when a node accepts a key agreed with a device off the body.

Alignment pairs each beat of node A with the nearest beat of node B within 300\,ms, so one node must publish its beat timestamps, which the processor receives in its peer list. These timestamps define the IBI sequence of the key, so an eavesdropper who receives them exactly reproduces $\mathbf{b}_A$ in every block. Waveform spoofing \cite{wang2025fakeppg} must reproduce the live IBI sequence at both sites within 300\,ms over a full window, while camera-based remote PPG measures heartbeat timing as accurately as contact PPG \cite{seepers2018rppg} and is the stronger threat.

\begin{table}[t]
\centering
\caption{Block Entropy, Residual Entropy After Helper Exposure, and Measured Key Recovery}
\label{tab:sec_leakage}
\scriptsize\setlength{\tabcolsep}{2pt}
\begin{tabular}{@{}lcc@{}}
\toprule
Estimate of $H(\mathbf{b}_A)$ & Bits/block & Residual \\
\midrule
Independent symbols, 1.88 to 1.97 bits & 120 to 126 & 34 to 40 \\
Shannon within window, $64\bar H_{\mathrm{w}}$ & 58.5 & 0.0 \\
Markov-1 rate, $64h_1$ & 72.6 & 0.0 \\
Markov-1 min-entropy, mean HR known & 3.8 & 0.0 \\
SP 800-90B, $64H_{\min}$ (final) & 0.2 & 0.0 \\
\midrule
Key recovered per block (\%) & H-W / H-A / W-A & All \\
\midrule
Node B, genuine & 70.6 / 50.5 / 55.2 & 57.3 \\
Eavesdropper, mean-HR guess & 50.1 / 40.1 / 49.7 & 45.2 \\
Node B, floor 1.0 bit & 46.8 / 35.3 / 35.6 & 38.3 \\
Eavesdropper, floor 1.0 bit & 10.2 / 9.0 / 8.0 & 9.1 \\
\bottomrule
\end{tabular}
\end{table}

\subsection{Entropy and Helper-Data Leakage}
\label{sec:sec_leak}

The pooled Shannon entropy of the Gray symbols is 1.882 to 1.974 bits per symbol. Within one window, however, the symbols concentrate in one or two bins and carry 0.816 to 0.940 bits per symbol. In 230 of the 2\,207 key blocks all 64 symbols are identical, and these blocks reconcile in 228 cases. Consecutive symbols are equal in 74.1\% of the cases. The NIST SP 800-90B tool \cite{turan2018sp80090b} rejects the assumption of independent and identically distributed symbols and returns 0.004 bits per symbol, against 1.606 bits for os.urandom data of equal length.

The helper is $\mathbf{h}=\mathbf{b}_A\oplus\mathbf{c}$, the polar transform is its own inverse over GF(2), and every codeword is zero on the frozen set $\mathcal{F}$ after the transform. Hence $u(\mathbf{h})_{\mathcal{F}}=u(\mathbf{b}_A)_{\mathcal{F}}$, and the eavesdropper reads 86 parity functions of $\mathbf{b}_A$. The residual min-entropy obeys $\tilde{H}_\infty(\mathbf{m}\mid\mathbf{h})\geq H_\infty(\mathbf{b}_A)-(N-K)$ \cite{dodis2008fuzzy}. Every dependence-aware estimate of $H(\mathbf{b}_A)$ lies below the 86 exposed bits (Table~\ref{tab:sec_leakage}), so the bound gives no guarantee. In Table~\ref{tab:sec_leakage}, $\bar H_{\mathrm{w}}$ is the mean within-window Shannon entropy of a symbol, $h_1$ the first-order Markov entropy rate and $H_{\min}$ (final) the final SP 800-90B estimate, each scaled to the 64 symbols of a block. The residual is the block entropy minus the 86 exposed bits, floored at zero. Only the min-entropy rows bound the message, and the Shannon and Markov rows serve as reference.

An eavesdropper who sets every symbol to the Gray symbol of the mean IBI and runs the decoder of node B recovers $\mathbf{m}$ in 45.2\% of the blocks, against 57.3\% for node B. The same guess recovers all three messages of 39.4\% of the session keys, and an order-aware Markov attack recovers $\mathbf{m}$ in 53.4\% of the blocks. An entropy floor at node A, which keeps a block only when its symbols carry at least 1.0 bit each, keeps 835 blocks, of which node B reconciles 38.3\% and the eavesdropper recovers 9.1\%. Timestamps rounded to 100, 250 and 500\,ms slots allow 72.0, 30.5 and 17.1\% recovery over all pairs. The biometric therefore does not protect the 42-bit message.

\subsection{Impostors and Quantizer Alternatives}
\label{sec:sec_quant}

 Impostors reach 54.7\% raw agreement on average, and 2.30 to 3.76\% of their attempts reconcile because windows of different people in the same bin give identical strings.

We tested four quantizers of the beat-to-beat variation (the IBI difference, the IBI minus a moving average, low-order IBI bits \cite{bao2008timing,zhang2012ipi}, and per-window quartiles). On WildPPG none of them reaches 40\% genuine agreement with a measured eavesdropper below 1\%. Within a window the IBI has a standard deviation of 46\,ms and the inter-site difference one of 43\,ms. Even with ideal reconciliation, this caps the shared secret per 64 beats at 15.3 bits in a Gaussian coherence estimate and at 20.9 and 24.9 bits in plug-in estimates for the two per-window quantizers \cite{maurer1993secret}.

\section{Limitations and Future Work}
\label{sec:discussion}
\label{sec:limits}

The front end, the foot detector and the SC decoder of HHK serve any IBI-based key scheme unchanged, so the hardware carries over to a secure variant of the protocol. Such a variant needs an alignment that publishes no beat timing, or a reconciliation code that tolerates inserted and deleted beats, together with more min-entropy per block than the 86 exposed bits. An entropy floor at node A is a first step in that direction, and order-aware attacks still recover 17.2\% of the blocks that pass it.

The board run covers 14 windows of 4 participants, while the agreement over all 16 participants comes from the fixed-point model that the board reproduces. Two boards that exchange their peer lists over a radio link would test the protocol end to end. The power figures come from Vivado on the routed design, and a current measurement on the board with a gated bus interface is the next step. For participant fex, the detected IBIs average 1\,185\,ms (50.6\,BPM) and 97\% of the key blocks consist of 64 identical Gray symbols, which may come from missed beats. A comparison with the ECG of this participant would identify the cause.

\section{Conclusion}
\label{sec:concl}

This paper has presented a processor for cross-location PPG key agreement in BANs. On 16 ambulatory participants the processor reaches 69.2\% key agreement between head and wrist, against 2.3\% for the best of five prior methods. On the FPGA the processor uses 8\,417 LUTs and 6\,259 FFs and consumes 261\,\textmu W. To our knowledge, the processor is the first synthesizable RTL architecture for IBI-based PPG key agreement in BANs evaluated on free-living ambulatory data across three body locations. It is also the first PPG key-agreement hardware reported with post-implementation resource and power figures.

The security evaluation quantifies the limits of absolute-IBI quantization with a fuzzy commitment on ambulatory data, where an eavesdropper who knows the mean heart rate recovers 39.4\% of the session keys. It also identifies the two requirements of a secure variant, an alignment that publishes no beat timing and more entropy per block. Because the front end, the foot detector and the SC decoder of HHK carry over unchanged, the processor provides the hardware basis for secure cross-location PPG key agreement in BANs.

\section*{Acknowledgment}
 The authors thank the AMD University Program for its support.

\bibliographystyle{IEEEtran}
\bibliography{references/references_tbiocas}

\end{document}